\documentclass[conference]{IEEEtran}
\IEEEoverridecommandlockouts
\usepackage{cite}
\usepackage{amsmath,amssymb,amsfonts}
\usepackage{graphicx}
\usepackage{textcomp}
\usepackage{xcolor}
\usepackage{cite}
\usepackage{balance}
\usepackage{bm}
\usepackage{subfigure}
\usepackage{algorithm} 
\usepackage{algorithmic}  
\usepackage[algo2e]{algorithm2e} 
\usepackage{savesym}
\usepackage{siunitx}
\def\BibTeX{{\rm B\kern-.05em{\sc i\kern-.025em b}\kern-.08em
    T\kern-.1667em\lower.7ex\hbox{E}\kern-.125emX}}

\DeclareMathOperator*{\argmax}{arg\,max}

\DeclareMathOperator{\EX}{\mathbb{E}}
\DeclareMathOperator{\diag}{diag} 
\begin{document}

\bstctlcite{IEEEexample:BSTcontrol}


\title{Spider RIS: Mobilizing Intelligent Surfaces for Enhanced Wireless Communications \\
\thanks{The work of Ibrahim Yildirim was supported by the Scientific and Technological Research Council of Turkey (TÜBİTAK) under Grant BİDEB 2214.}
\thanks{The work of Ertugrul Basar was also supported by TÜBİTAK under Grants 120E401.}
\thanks{The work of Tho Le-Ngoc was also supported in part by the Natural Sciences and Engineering Research Council of Canada (NSERC), InterDigital Canada, and Prompt Quebec under an NSERC Alliance Grant.}
}

\author{\IEEEauthorblockN{Ibrahim Yildirim\textsuperscript{$\ast$,$\bullet$,$\circ$}, Mobeen Mahmood\textsuperscript{$\circ$}, Ertugrul Basar\textsuperscript{$\ast$}, Tho Le-Ngoc\textsuperscript{$\circ$}}
	\IEEEauthorblockA{\textsuperscript{$\ast$}CoreLab, Department of Electrical and Electronics Engineering, Koç University, Sariyer 34450, Istanbul, Turkey \\
\textsuperscript{$\bullet$}Faculty of Electrical and Electronics Engineering, Istanbul Technical University, Sariyer 34469, Istanbul, Turkey.\\
		\textsuperscript{$\circ$}Department of Electrical and Computer Engineering, McGill University, Montreal, QC, Canada \\
		Email: yildirimib@itu.edu.tr, mobeen.mahmood@mail.mcgill.ca, ebasar@ku.edu.tr,
		tho.le-ngoc@mcgill.ca
		\vspace{-3ex}}
}

\maketitle

\begin{abstract}
In this study, we introduce Spider RIS technology, which offers an innovative solution to the challenges encountered in movable antennas (MAs) and unmanned aerial vehicle (UAV)-enabled communication systems. By combining the dynamic adaptation capability of MAs and the flexible location advantages of UAVs, this technology offers a dynamic and movable RIS, which can flexibly optimize physical locations within the two-dimensional movement platform. Spider RIS aims to enhance the communication efficiency and reliability of wireless networks, particularly in obstructive environments, by elevating the signal quality and achievable rate.
The motivation of Spider RIS is based on the ability to fully exploit the spatial variability of wireless channels and maximize channel capacity even with a limited number of reflecting elements by overcoming the limitations of traditional fixed RIS and energy-intensive UAV systems. 
 Considering the geometry-based millimeter wave channel model, we present the design of a three-stage angular-based hybrid beamforming system empowered by Spider RIS:  First, analog beamformers are designed using angular information, followed by the generation of digital precoder/combiner based on the effective channel observed from baseband stage. Subsequently, the joint dynamic positioning with phase shift design of the Spider RIS is optimized using particle swarm optimization, maximizing the achievable rate of the systems.
\end{abstract}


\section{Introduction}


The search for transformative technologies to meet evolving connectivity demands remains paramount in the relentless drive toward the emergence of 6G and the next generation of wireless networks. Key applications of the 6G wireless network include providing users with immersive experiences such as high-definition virtual reality, mobile holography, and digital twins \cite{6G_Guan}. These applications necessitate exceptionally high data rates and massive connectivity, concurrently requiring high reliability and low latency that promises performance close to wired communication standards \cite{6G_Dang}. To achieve these goals, 6G is expected to embrace innovative technologies that dynamically program changing wireless signal propagation environments, such as reconfigurable intelligent surfaces (RISs). 
RISs stand out as a transformative tool for wireless networks to achieve various network goals such as coverage extension, environmental sensing, positioning, and spatiotemporal focusing.
Composed of controllable arrays of low-cost passive elements, RISs act as intelligent mediators that modify the reflections of the incoming signals. These intelligent and cost-effective capabilities of the RISs are expected to contribute considerably to realizing the 6G vision by providing a programmable environment to meet the stringent requirements and anticipated use cases of next-generation wireless networks \cite{RIS_Survey_Jian,SimRIS_Mag,Basar_Access_2019, Yildirim_multiRIS}.


On the other hand, today's rapidly evolving wireless communications needs shed significant light on the capabilities of unmanned aerial vehicles (UAVs) \cite{UAV_SURVEY,UAV_RIS,UAV_RIS_Pogaku}. While UAVs offer 3D mobility, they encounter distinct shortcomings in positioning and energy consumption. High energy consumption is a severe problem in applications requiring long-distance and persistent communication. UAVs require battery replacement or recharging to provide continuous communication services, which limits their flight time and operational range \cite{UAV_Zhou}. Additionally, using UAVs in an indoor environment poses severe practical difficulties. Although capable of mobility, UAVs are challenging to maneuver indoors, which can affect the coverage and reliability of transmission. From these perspectives, addressing these battery and maneuver-related challenges is essential for successfully deploying and operating UAV-based communication systems \cite{UAV_Li}.

In current multiple-input multiple-output (MIMO) systems, antennas are deployed in fixed positions, which prevents taking full advantage of the degrees of freedom (DoF) in the continuous spatial domain to optimize spatial multiplexing performance. To overcome this fundamental limitation, the movable antenna (MA) system has been recently proposed as a new solution to fully evaluate wireless channel variations in a continuous spatial domain \cite{Movable_Ant_2023_2}. Unlike fixed-position antennas, each MA is connected to the radio frequency (RF) chain via a flexible cable, allowing its position to be flexibly adjusted in a given spatial region to achieve more favorable channels to improve communication performance. In MA-supported MIMO systems, channel capacity is maximized by simultaneously adjusting the positions of MAs in the areas where the transmitter and receiver are located \cite{Movable_Ant_2023_1}. In MA systems, the need for all antenna elements to change position independently and to provide perfect channel state information for each position makes it challenging to use these systems for massive MIMO applications. 
Optimization of such systems increases algorithmic and computational complexity, making it complicated in terms of cost and practicality to implement for large-scale massive MIMO systems.

\begin{figure}[!t]
	\centering
	\includegraphics[width=0.8\columnwidth]{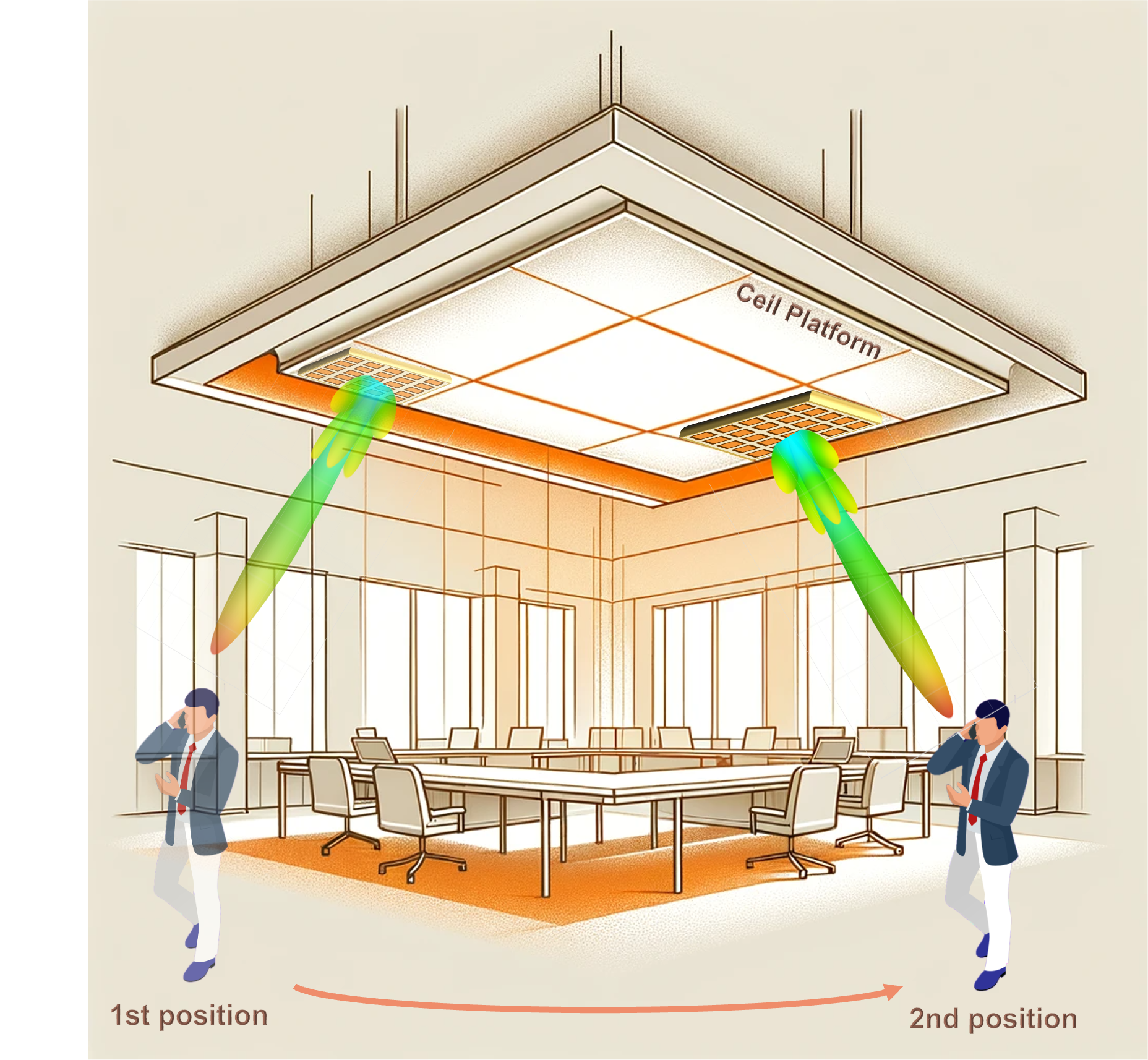}
	\vspace{-2ex}
	\caption{Illustration of a moving \textit{Spider RIS } hanging on the ceiling platform in an indoor office environment according to changing positions.}
	\vspace{-2ex}
	\label{fig_ceil} 
\end{figure}

In this study, we propose the movable RIS concept, called "Spider RIS", which is an innovative system in which the RIS is positioned on a wall-mounted platform and aims to increase the degrees of freedom and performance of the communication system by moving in a limited space as illustrated in Fig. \ref{fig_ceil}. While each antenna element moves independently in the MA systems, in Spider RIS, the entire RIS moves together with the all of its elements. This allows the RIS to be shifted horizontally or vertically within a given spatial region, changing its position to find optimal signal path or channel conditions and thus improve transmission quality. The movement of the Spider RIS can be controlled through motors or other mechanical systems integrated into the platform. This design eliminates the need to control individual antenna elements separately, thereby reducing system complexity and cost while maintaining the ability to reshape the channel matrix flexibly. With these features, Spider RIS has the potential to provide higher efficiency and performance by overcoming the limitations of existing RIS technologies in wireless communication systems. An alternative implementation of the Spider RIS may involve designating the entire movement area on the ceiling or a horizontal wall as the RIS, dynamically activating specific sections for transmission. However, this design becomes impractical in larger indoor or outdoor environments when considering the cost of channel estimation and covering the entire surface with RIS.

Spider RIS is a technology inspired by both MAs and UAVs-supported communication systems while aiming to overcome the challenges brought by these two technologies. Combining the improved spatial DoF capability offered by MAs and the flexible positioning advantages of UAVs, it offers a dynamic and movable RIS capable of dynamically optimizing the physical positions within the movement platform. 
The motivation of Spider RIS is based on the ability to fully exploit the spatial variability of wireless channels and maximize channel capacity even with a limited number of reflecting elements. This approach aims to overcome the limitations that traditional fixed RIS and energy-intensive UAV systems cannot overcome and to open new horizons in wireless communications.

Considering the above advantages of Spider RIS, an angular-based hybrid beamforming (AB-HBF) system is designed in this study for millimeter wave (mmWave) massive MIMO (mMIMO) systems, which is based on the slowly time-varying angular parameters of the channel and aims to minimize the instantaneous channel estimation overhead \cite{ASIL_ABHP_Access, RIS_AB_HPC}. In particular, the RF stage is built on slowly time-varying angular information, and the BB stage is built on reduced-size channel state information (CSI). Considering the mmWave channel model based on 3D geometry, a three-stage HBF  design for the Spider RIS-aided mMIMO system is proposed: (i) RF beamformers, (ii) baseband (BB) precoder/combiner, and (iii) joint dynamic RIS Positioning and phase shift design for Spider RIS. First, transmit/receive RF beamformers are designed using the slowly time-varying departure/arrival angle information (AoD/AoA) of the channel, and then the BB precoder and combiner are constructed using the effective channel seen from the BB stage. Then, the RIS phase shift matrix and trajectory of the RIS are designed to maximize the achievable rate of massive MIMO systems through a nature-inspired particle swarm optimization (PSO) algorithm.
\section{System and Channel Model of Spider RIS}
\vspace{-1ex}
In this section, we introduce the system and channel models of the proposed Spider RIS deployment in mMIMO systems.
\vspace{-1ex}
\subsection{System Model}
We consider a point-to-point mmWave mMIMO indoor environment, where the direct line-of-sight (LoS) communication paths between the transmitter (Tx) and receiver (Rx) are frequently obstructed by various obstacles, such as walls, furniture, and other indoor structures. A movable RIS is placed on the ceiling of the floor to address the following challenges: 1) Overcome obstructions; 2) dynamic adoption\footnote{The movable RIS can adapt to the ever-changing indoor environment, and it can adjust in real-time to the movements of users.}; and 3) enhance spectral efficiency. Unlike static RIS, which is deployed at a fixed location, we introduce the Spider RIS to study the performance gains in hybrid mMIMO systems. \par 

\begin{figure*}[!t]
	\includegraphics[height= 4cm, width=1.9\columnwidth]{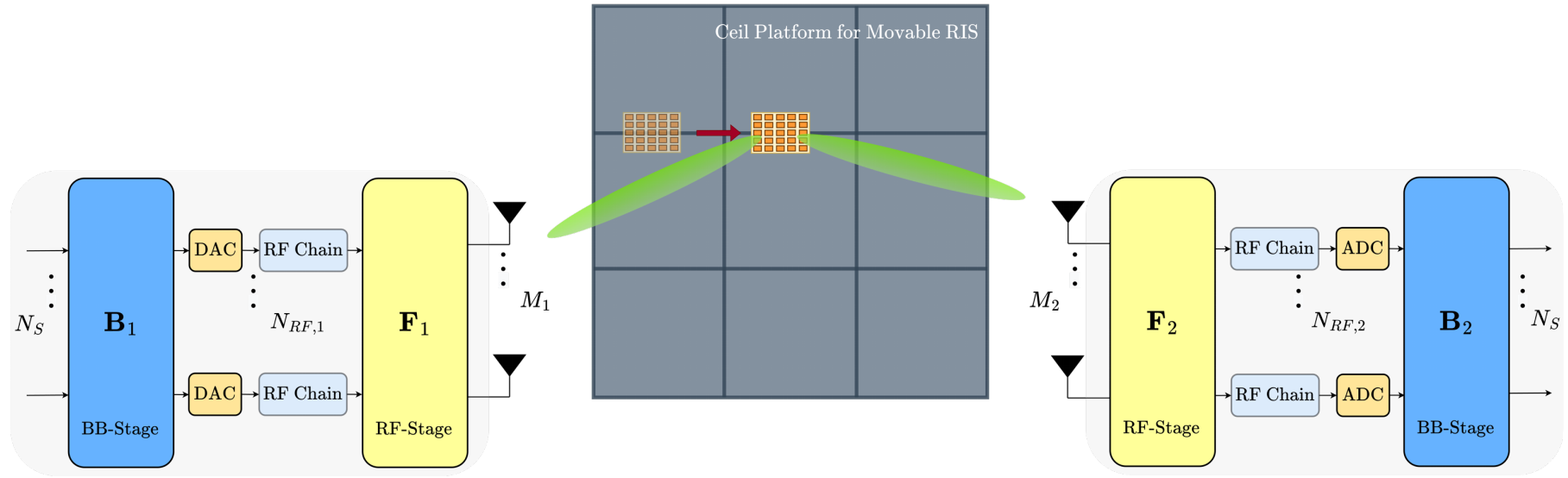} 
	\vspace{-1em}
	\caption{Movable RIS-assisted mMIMO HBF system model.}
	\label{fig:fig1}
	\vspace{-3ex}
\end{figure*} 
As shown in Fig. \ref{fig:fig1}, we consider a BS to be equipped with $M_1 = M_{1_x} \times M_{1_y}$ antennas, a Rx with $M_2 = M_{2_x} \times M_{2_y}$ antennas, and a Spider RIS with $M_I = M_{I_x} \times M_{I_y}$ dynamic reflecting elements. The location of Tx, RIS, and Rx is denoted by $\left(x_1, y_1, z_1\right)$, $\left(x_r, y_r, z_r\right)$ and $\left(x_2, y_2, z_2 \right)$, respectively. We assume a uniform rectangular array (URA) configuration for the Tx, RIS, and Rx such that $M_{i_x}$ and $M_{i_y}$ represent the number of antenna/reflector elements along $x$ and $y$-axis, respectively, where $i \in \{1,2, I\}$. We consider HBF architecture, where the Tx consists of an RF beamforming stage $\mathbf{F}_1$ $\in$ $\mathbb{C}^{M_1 \times N_{{RF}_1}}$ and BB stage $\mathbf{B}_1$ $\mathbb{C}^{N_{{RF}_1} \times N_S}$. Here, $N_S$ represents the number of data streams from Tx to Rx through channel $\mathbf{H}$ $\in$ $\mathbb{C}^{M_{2} \times M_1}$, and $N_{{RF}_1}$ is the number of RF chains such that $N_S \leq N_{{RF}_1} \leq M_1$ to guarantee multi-stream transmission. Considering the transmitted signal is $\mathbf{d} = [d_1, d_2, \hdots, d_{N_S}]^T$ with $\EX \{\mathbf{d}\mathbf{d}^H\} = \mathbf{I}_{N_S}$ $\in$ $\mathbb{C}^{N_S \times N_S}$, then the signal transmitted by Tx is given as follows: \vspace{-1ex}
\begin{equation}
	\mathbf{s} = \mathbf{F}_1\mathbf{B}_1 \mathbf{d}. \label{eq:1} \vspace{-1ex}
\end{equation}
The power constraint of the beamforming matrices can be expressed as $\lVert \mathbf{F}_1\mathbf{B}_1 \rVert^2_F$ = $P_T$, where $P_T$ denotes the total transmit power. Due to unfavorable propagation conditions (i.e., severe blockage) of the direct link, the transmitted signal arrives at Rx via the Tx-RIS-Rx channel. Let $\mathbf{H}_{TI} \in \mathbb{C}^{M_I \times M_1}$ denote the channel from the Tx to RIS, and $\mathbf{H}_{IR}$ $\in$ $\mathbb{C}^{M_2 \times M_I}$ is the RIS-Rx channel. Then, the received signal at the Rx via Spider RIS is
given as follows:
\begin{equation}
	\begin{split}
		\mathbf{y} &= \mathbf{H} \mathbf{s} + \mathbf{n}  = (\mathbf{H}_{IR}\mathbf{\Phi}\mathbf{H}_{TI})\mathbf{s}  + \mathbf{n}, \\
  &=(\mathbf{H}_{IR}\mathbf{\Phi}\mathbf{H}_{TI})\mathbf{F}_1\mathbf{B}_1 \mathbf{d}  + \mathbf{n},
	\end{split}	\label{eq:2}  \vspace{-1em}
\end{equation}
where $\mathbf{n}$ $\sim$ $\mathcal{CN} (\mathbf{0}, \sigma_n^2 \mathbf{I}_{M_2})$ represents the additive white Gaussian noise at the Rx and $\mathbf{\Phi} \triangleq \diag(e^{j\phi_1}, \cdots, e^{j\phi_{M_I}})$ $\in \mathbb{C}^{M_I \times M_I}$ is the RIS phase matrix and $\phi_i \in [0, 2\pi]$ denote the phase shift introduced by the $i^{th}$ element of the RIS. At the Rx, the signal received is processed through RF stage $\mathbf{F}_2$ $\in$ $\mathbb{C}^{N_{{RF}_2} \times M_2}$ and BB stage $\mathbf{B}_2 \in \mathbb{C}^{N_S \times N_{{RF}_2}}$. Then, the received signal at the Rx after combining can be written as: 
\begin{equation}
\tilde{\mathbf{y}} = \mathbf{B}_2 \mathbf{\mathcal{H}}\mathbf{B}_1 \mathbf{d} + \mathbf{B}_2\mathbf{F}_2\mathbf{n} \label{eq:3} \vspace{-0.5ex}
\end{equation}
where $\bf{\mathcal{H}} = \mathbf{F}_{2}\mathbf{H}\mathbf{F}_1$ 
is the effective channel matrix seen from the BB stages. For a movable RIS-aided hybrid mMIMO system, where the RIS is positioned at a fixed height $z_r$ (i.e., ceiling), the total achievable rate can be maximized by the joint optimization of RF beamforming/combining and BB stages $\mathbf{F}_1$, $\mathbf{B}_1$, $\mathbf{F}_2$, $\mathbf{B}_2$, the phase shifts at RIS $\mathbf{\Phi}$ as well as the RIS position ($x_r$, $y_r$) within a given deployment area. Then, we can formulate the optimization problem as follows:
\begin{equation}
	\begin{split}
		&\max_{\left\{ \mathbf{F}_1, \mathbf{B}_1, \mathbf{F}_2, \mathbf{B}_2, \mathbf{\Phi}, \mathbf{x}_o \right\}} \hspace{-2ex} \quad \log_2 \left|\mathbf{I}_{N_S}  + \mathbf{W}_{\tilde{n}}^\mathrm{-1}\mathbf{B}_2 \mathbf{\mathcal{H}} \mathbf{B}_1 \mathbf{B}_1^H \mathbf{\mathcal{H}}^H \mathbf{B}_2^H \right| \\
		 &\textrm{s.t.} \hspace{2ex} C_1: \hspace{1ex}  |\mathbf{F}_{1} (i,j)| = \frac{1}{\sqrt{M_1}}, |\mathbf{F}_{2} (i,j)| = \frac{1}{\sqrt{M_2}} \hspace{1ex} \forall i,j, \\
		 & \quad \quad C_2:  \hspace{1ex} \EX \{ \left\lVert \mathbf{s} \right\rVert_2^2\} \leq P_T, \\ 
		&\quad \quad C_3: \hspace{1ex} \mathbf{\Phi} = \diag(e^{j\phi_1}, \cdots, e^{j\phi_{M_I}}),  \\
        &\quad \quad C_4: \hspace{1ex} \phi_i \in [0, 2\pi], \hspace{1ex} \forall i,  \\
		&\quad \quad C_5: \hspace{1ex} x_{\mathrm{min}} \leq \mathbf{x}_o \leq x_{\mathrm{max}}, 
	\end{split} \label{eq:optimization}  \raisetag{0.8\baselineskip} 
\end{equation} 
where $\mathbf{W}_{\tilde{n}} = \sigma_n^2 \mathbf{B}_2 \mathbf{F}_2\mathbf{F}_2^H \mathbf{B}_2^H$ is the covariance matrix of the noise $\tilde{n} = \mathbf{B}_2\mathbf{F}_2\mathbf{n}$. Here, $C_1$ refers to the constant-modulus (CM) constraint due to the use of phase shifters at Tx and Rx, $C_2$ indicates the total transmit power constraint, $C_3$ specifies the phase matrix design to shape signal propagation, $C_4$ limits the phase values of RIS elements within [0, 2$\pi$], and $C_5$ establishes the RIS deployment boundary within the indoor environment. The optimization problem defined in (\ref{eq:optimization}) is challenging due to the non-convexity of the objective function and the CM constraint due to RF beamforming/combining stages. To solve this problem, we first develop HBF stages for the Tx and Rx based on arbitrary RIS location to optimize $\mathbf{x}$. Then, based on the optimal RIS location and phase matrix $\mathbf{\Phi}$, we sequentially update the RF and BB stages.
\vspace{-1ex}
\subsection{Channel Model}
We consider mmWave channels for both Tx-RIS and RIS-Rx links. While LoS channel models can be useful for simple scenarios, they can be limited in their ability to capture the channel complexities (e.g., multi-path fading and shadowing), especially in an indoor environment. Thus, mmWave channel models can provide a more accurate representation of the channel characteristics, including the impact of non-LoS paths and obstacles on the signal propagation in RIS-assisted mMIMO communications. Based on the Saleh-Valenzuela channel model \cite{ju2021millimeter}, the channel is given as follows: \vspace{-1ex}
\begin{equation}
	\mathbf{H}_i = \sum\nolimits_{l = 1}^{L} \frac{z_{l,i}}{\alpha \tau_{i}^\eta}  \mathbf{a}_r \big( {{\theta_{l,i}^{(r)}}},{{\psi_{l,i}^{(r)}}} \big) \mathbf{a}_t^T \big( {{\theta_{l,i}^{(t)}}},{{\psi_{l,i}^{(t)}}} \big), \label{eq:channel_vector} \vspace{-1ex}
\end{equation}
where $L$ is the total number of paths, $\alpha = 32.4 + 20 \log_{10}(f_c)$ represents the reference path loss in dB, $\eta$ is the path loss exponent, and $z_{l,i}$ is the complex gain of $l^{th}$ path in $i^{th}$ channel, where $i \in \{IR, TI\}$, and $\bf{a}( .\hspace{0.25ex}, \hspace{0.25ex}. )$ is the Tx/Rx array steering vector for URA, which is given as \cite{mobeen_UAV_DL}: \vspace{-0.5ex}
\begin{equation}\label{eq_phase_vector}
	\begin{split}
		{\bf{a}}_j\hspace{-0.5ex}\left( {{\theta, \psi}} \right) \hspace{-0.5ex}&=\hspace{-0.75ex} \big[ {1,{e^{- j2\pi d  {{\sin (\theta) \cos (\psi)}} }}, \cdots,{e^{ -j2\pi d\left( {M_x - 1} \right) {{\sin (\theta) \cos (\psi)}} }}} \big]\\
	& \hspace{-2ex}\otimes \hspace{-0.5ex}\big[ {1,{e^{ -j2\pi d  {{\sin (\theta) \sin (\psi)}} }}\hspace{-0.25ex}, \hspace{-0.25ex}\cdots\hspace{-0.25ex},\hspace{-0.25ex}{e^{ -j2\pi d\left( {M_y - 1} \right) {{\sin (\theta) \sin (\psi)}} }}} \big]\hspace{-0.25ex}, \hspace{-2ex}
	\end{split} \vspace{-0.5ex} \raisetag{0.8\baselineskip}
 \vspace{-1em}
\end{equation}
where $j = \{t,r\}$, $M$ is the corresponding Tx/Rx antennas, $d$ is the inter-element spacing, and $M_x(M_y)$ is the horizontal (vertical) size of the array. Here, the angles ${\theta _{l,i}^{(t)}} \in \big[ {{\theta _{i}^{(t)}} - {\delta_{i}^\theta}}, {{\theta _{i}^{(t)}} + {\delta _{i}^\theta}} \big]$ and   
${\psi _{l,i}^{(t)}} \in \big[ {{\psi _{i}^{(t)}} - {\delta _{i}^\psi}}, {{\psi _{i}^{(t)}} + {\delta _{i}^\psi}} \big]$ are the elevation AoD (EAoD) and azimuth AoD (AAoD) for $l^{th}$ path in $i^{th}$ channel, respectively. Similarly, the angles ${\theta _{l,i}^{(r)}} \in \big[ {{\theta _{i}^{(r)}} - {\delta_{i}^\theta}}, {{\theta _{i}^{(r)}} + {\delta _{i}^\theta}} \big]$ and ${\psi _{l,i}^{(r)}} \in \big[ {{\psi _{i}} - {\delta _{i}^\psi}}, {{\psi _{i}^{(r)}} + {\delta _{i}^\psi}} \big]$ are the elevation AoA (EAoA) and azimuth AoA (AAoA). $\theta_{i}(\psi_{i})$ is the mean EAoD(AAoD)/EAoA(AAoA) with angular spread $\delta_{i}^\theta (\delta_{i}^\psi)$.  
\vspace{-1ex}
\section{Joint Hybrid Beamforming, RIS Dynamic Positioning and Phase Shift Design}\vspace{-2ex}
In this section, our objective is to optimize RIS deployment for a dynamic indoor environment jointly with its phase shift design and sequentially construct HBF stages for the Tx and Rx to reduce the CSI overhead size while maximizing the throughput of a movable RIS-assisted mMIMO system. First, we discuss the design of the RF beamformer/combiner and BB stages, which is followed by the PSO-based algorithmic solution to jointly optimize the RIS location and its phase shifts to enhance the capacity. \vspace{-1ex}
\subsection{RF Beamforming/Combining Stage Design}
The RF beamforming stage for the Tx and Rx are designed as follows:\vspace{-2ex}
\begin{equation}\label{eq:eq_TX_RF_1}
	{{\bf{F}}}_1 \hspace{-0.45ex}= \hspace{-0.45ex}\big[ \hspace{-0.25ex}{\bf{e}}_t\hspace{-0.25ex}\big( {\lambda ^{u_1\hspace{-0.15ex}}_{x}\hspace{-0.25ex},\lambda ^{k_1\hspace{-0.15ex}}_{y}} \big)\hspace{-0.25ex}, \hspace{-0.15ex}\cdots\hspace{-0.35ex},\hspace{-0.15ex} {\bf{e}}_t\big( \hspace{-0.25ex}{\lambda ^{u_{N_{{RF}_1}\hspace{-0.15ex}}\hspace{-0.15ex}}_{x}\hspace{-0.15ex},\lambda ^{k_{N_{{RF}_1}}}_{y}} \hspace{-0.25ex}\big) \hspace{-0.25ex}\big] \hspace{-0.75ex}\in \hspace{-0.5ex}\mathbb{C}^{M_{1} \hspace{-0.2ex}\times N_{{RF}_1}},
\end{equation}
\begin{equation}\label{eq:eq_RX_RF_1}
{{\bf{F}}_2} \hspace{-0.75ex}=\hspace{-0.75ex} \big[ \hspace{-0.25ex}{{\bf{e}_r}\big(\hspace{-0.35ex} {\lambda ^{u_1}_{x} \hspace{-0.35ex},\lambda ^{k_1}_{y}\hspace{-0.15ex}} \hspace{-0.5ex}\big), \hspace{-0.35ex} \cdots\hspace{-0.25ex},\hspace{-0.35ex} {\bf{e}}_r\big(\hspace{-0.25ex} {\lambda ^{u_{N_{{RF,2}}}\hspace{-0.15ex}}_{x},\lambda ^{k_{N_{{RF,2}}}\hspace{-0.15ex}}_{y}} \hspace{-0.35ex}\big)}\hspace{-0.25ex} \big]^T\hspace{-0.75ex} \hspace{-0.95ex} \in \hspace{-0.5ex}\mathbb{C}^{N_{{RF,2\hspace{-0.15ex}}\hspace{-0.15ex}} \times M_2},\hspace{-0.25ex} 
\end{equation}
where ${\bf{e}}_j(\cdot,\cdot)$ is the corresponding transmit or receive steering vector, which is defined as: ${\bf{e}}_j\hspace{-0.5ex}\left( {{\theta, \psi}} \right) \hspace{-0.5ex} =\hspace{-0.75ex} \frac{1}{\mathcal{M}}\big[ {1,{e^{j2\pi d  {{\sin (\theta) \cos (\psi)}} }}, \cdots,{e^{j2\pi d\left( {\mathcal{M}_{x} - 1} \right) {{\sin (\theta) \cos (\psi)}} }}} \big]^T \otimes \hspace{-0.5ex}\big[ {1,{e^{j2\pi d  {{\sin (\theta) \sin (\psi)}} }}, \cdots,{e^{j2\pi d\left( {\mathcal{M}_{y} - 1} \right) {{\sin (\theta) \sin (\psi)}} }}} \big]^T$, where $\mathcal{M} = \{M_1,M_2\}$ for $j = \{t,r\}$. Here, the RF beamformers are constructed via quantized angle-pairs having a boundary, which are defined as ${{\lambda _{x}^u} = -1 + \frac{2u-1}{{{\mathcal{M}_{x}}}}}$ for $u = 1, \cdots, {\mathcal{M}_{x}}$ and ${{\lambda _{y}^k} = -1 + \frac{2k-1}{{{\mathcal{M}_{y}}}}}$ for $k = 1, \cdots, {\mathcal{M}_{y}}$. The quantized angle-pairs reduce the number of RF chains while providing complete AoD/AoA support defined as:
\begin{align}
	\textrm{AoD} &= \left\lbrace \sin \left( \theta \right)\left[ \cos \left( \psi \right),\sin \left( \psi \right) \right] \Big| \theta \in \bm{\theta}_t, \psi \in \bm{\psi}_t \right\rbrace, \label{eq_AoD_Supp} \\
	\textrm{AoA} &= \left\lbrace \sin \left( \theta \right)\left[ \cos \left( \psi \right),\sin \left( \psi \right) \right] \Big| \theta \in \bm{\theta}_r, \psi \in \bm{\psi}_r \right\rbrace, \label{eq_AoA_Supp}
\end{align}
where $\bm{\theta}_i = \left[ {\theta_i - \delta_i^\theta}, {\theta_i + \delta_i^\theta} \right]$ and $\bm{\psi}_i = \left[ {\psi_i - \delta_i^\psi}, {\psi_i + \delta_i^\psi} \right]$ denote the azimuth and elevation angle supports. \vspace{-2ex} 
\subsection{BB Precoder/Combiner Design}\vspace{-1ex} 
After designing the RF stages, the effective channel matrix $\bf{\mathcal{H}}$ as seen from the BB stage is given as:
\begin{equation} 
	\bf{\mathcal{H}} = \mathbf{F}_{2}\mathbf{H}\mathbf{F}_1 = \mathbf{U} \mathbf{\Sigma} \mathbf{V}^{H},
\end{equation}
where $\mathbf{U} \in \mathbb{C}^{N_{{RF}_2} \times \text{rank}(\bf{\mathcal{H}})}$ and $\mathbf{V} \in \mathbb{C}^{N_{{RF}_1} \times \text{rank}(\bf{\mathcal{H}})}$ are tall unitary matrices. Here, $\mathbf{\Sigma}$ is the diagonal matrix with singular values in decreasing order. Assuming rank$(\bf{\mathcal{H}}) \geq N_S$, $\mathbf{V}$ can be partitioned as $\mathbf{V} = [\mathbf{V}_{1}, \mathbf{V}_{2}]$ with $\mathbf{V}_{1}$ $\in$ $\mathbb{C}^{\mathcal{N_{{RF}_1}} \times N_S}$.
Then, the optimal $\mathbf{B}_1$ and $\mathbf{B}_2$ can be obtained as \cite{mobeen_spherical_UAV}:
\begin{equation}
\mathbf{B}_1 = \sqrt{\frac{P_T}{N_S}}\mathbf{V}, \hspace{3ex} \mathbf{B}_2 = \mathbf{U}^H. \label{eq:BB_1st_link}
\end{equation} 
\subsection{Joint Dynamic RIS Positioning \& Phase Shift Design}
After the design of RF and BB stages for the Tx and Rx, the optimization problem given in (\ref{eq:optimization}) can be reformulated as:  
\begin{equation}
	\begin{aligned}
		&\max_{\left\{\mathbf{\Phi}, \mathbf{x}_o \right\}} \hspace{-2ex} \quad \mathrm{R}_T = \log_2 \left|\mathbf{I}_{N_S}  + \mathbf{W}_{\tilde{n}}^\mathrm{-1}\mathbf{B}_2 \mathbf{\mathcal{H}} \mathbf{B}_1 \mathbf{B}_1^H \mathbf{\mathcal{H}}^H \mathbf{B}_2^H \right| \\
		& \hspace{5ex}\textrm{s.t.} \hspace{5ex} C_3 - C_5. 
	\end{aligned} \label{eq:optimization_reformulated} 
\end{equation} 

\makeatletter
\newcommand\fs@betterruled{%
\def\@fs@cfont{\bfseries}\let\@fs@capt\floatc@ruled
  \def\@fs@pre{\vspace*{5pt}\hrule height.8pt depth0pt \kern2pt}%
  \def\@fs@post{\kern2pt\hrule\relax}%
  \def\@fs@mid{\kern2pt\hrule\kern2pt}%
  \let\@fs@iftopcapt\iftrue}
\floatstyle{betterruled}
\restylefloat{algorithm}
\makeatother
\begin{algorithm}
[t!]\label{algo:3}
	\textbf{Input}:  $Z$, $T$, ($\theta$, $\psi$), ($x_1$,$y_1$,$z_1$), ($x_r$,$y_r$,$z_r$), ($x_2$,$y_2$,$z_2$). \\
	 \textbf{Output}: $\mathbf{x}_o$, $\mathbf{P}_{\Phi}$, $\mathbf{F}_{1}$, $\mathbf{B}_{1}$, $\mathbf{F}_{2}$, $\mathbf{B}_{2}$. \\
	\SetAlgoLined 
	Formulate  RF and BB stages at the Tx using (\ref{eq:eq_TX_RF_1}), (\ref{eq:BB_1st_link}) \\
	Formulate  HBF stages at the Rx using (\ref{eq:eq_RX_RF_1}), (\ref{eq:BB_1st_link}) \\
	\For{$z = 1:Z$}{
		Initialize the velocity as $\mathbf{J}_{v,z}^{(0)} = \bf{0}$. \\
		Each entry of $\mathbf{J}_{p,z}^{(0)}$ \hspace{-1ex}is uniformly distributed in $[0,1]$.\\
		Set the personal best $\mathbf{J}_{p,\mathrm{best},z}^{(0)} = \mathbf{J}_{p,z}^{(0)}$.
	}
	Find the global best $\mathbf{J}_{p,\mathrm{best}}^{(0)}$ as in (\ref{eq: global_best_PSOLPA}).\\
	\For {$t = 1:T$}{
		\For{$z = 1:Z$}
		{
			Update the velocity $\mathbf{J}_{v,z}^{(t)}$ as in (\ref{eq:velocity_PSOLPA}).\\
			Update the position $\mathbf{J}_{p,z}^{(t)}$ as in (\ref{eq:position_PSOLPA}).\\
			Find the personal best $\mathbf{J}_{p,\mathrm{best},z}^{(t)}$ as in (\ref{eq: personal_best_PSOLPA}).}
		Find the global best $\mathbf{J}_{p,\mathrm{best}}^{(t)}$ as in (\ref{eq: global_best_PSOLPA}).
	}
	$\mathbf{x}_o = \mathbf{X}_{\mathrm{best}}^{(T)}$, $\mathbf{\Phi} = \diag(e^{(j\mathbf{P}_{\Phi_{\mathrm{best}}})})$ \\
	Update $\mathbf{B}_1, \mathbf{B}_{2}$ for $\mathbf{x}_o$ and $\mathbf{\Phi}$.
	\caption{Proposed Joint HBF, RIS Phase Shift and Deployment Algorithm} 
\end{algorithm}
This resulting problem in (\ref{eq:optimization_reformulated}) is intractable due to the joint dependence of both the RIS phase value $\phi_i \in [0, 2\pi]$, $\forall i = \{1, \cdots, M_I\}$ and the RIS deployment $\mathbf{x} = [x_o, y_o]^T$ on the rate expression in (\ref{eq:optimization}). To overcome this challenge, we propose sequential optimization using swarm intelligence, which employs multiple agents, called particles, to explore the search space of the objective function given in (\ref{eq:optimization_reformulated}). First, a total of $Z$ particles are randomly placed in search space such that each particle communicates with one another to share their personal best and update the current global best solution for the objective function. The particles then move iteratively for a total of $T$ iterations to reach the global optimum solution. In particular, each RIS location $\mathbf{x} = [x_r, y_r]^T$ and $\mathbf{\Phi}$ are optimized by using a PSO-based algorithmic solution while maximizing the total achievable rate of RIS-assisted mMIMO system. Particularly, each $z^{th}$ particle at the $t^{th}$ iteration now represents an instance of each RIS location and its phase value for a total of $M_I$ RIS elements, which is given as follows:
\begin{equation}
	\begin{split} \hspace{-1ex}
\mathbf{J}_{\hspace{-0.25ex}z\hspace{-0.25ex}}^{\hspace{-0.25ex}(t)\hspace{-0.25ex}} \hspace{-0.35ex}=\hspace{-0.35ex} [\hspace{-0.25ex}\mathbf{X}_z^{\hspace{-0.25ex}(t)}\hspace{-0.5ex}, \mathbf{P}_{\hspace{-0.25ex}\Phi_z}^{(t)}\hspace{-0.25ex}]^T \hspace{-0.95ex}=\hspace{-0.5ex} [x_{z}^{(t)\hspace{-0.25ex}}\hspace{-0.25ex}, y_{z}^{(t)\hspace{-0.25ex}}\hspace{-0.25ex},p_{{\phi_{1,z}}\hspace{-0.25ex}}^{(t)}\hspace{-0.5ex},\cdots,\hspace{-0.25ex} p_{{\phi_{M_I,z}}\hspace{-0.25ex}}^{(t)}]^T \hspace{-0.75ex} \in\hspace{-0.55ex} \mathbb{R}^{\hspace{-0.25ex}M_I + 2\hspace{-0.25ex}}, 
	\end{split}    \label{eq:Joint_PSO}    \raisetag{1\baselineskip}
\end{equation}
where each $z^{th}$ particle represents the RIS position and its phase value and calculates the objective function as $R_{\hspace{-0.15ex}T}(\hspace{-0.15ex}\mathbf{F}_1\hspace{-0.15ex}, \mathbf{B}_1\hspace{-0.15ex}, \mathbf{F}_{\hspace{-0.15ex}2\hspace{-0.15ex}}, \mathbf{B}_{2}, \mathbf{X}_z^{(t)}\hspace{-0.15ex}, \mathbf{P}_{\Phi_z}^{(t)})$. The position $\mathbf{J}_{p_z}^{(t)}$ and velocity $\mathbf{J}_{v_z}^{(t)}$ for $z^{th}$ particle during $t^{th}$ iteration are updated as follows:
\begin{equation}
	\mathbf{J}_{p_z}^{(t+1)}= \mathbf{J}_{p_z}^{(t)} + \mathbf{J}_{v_z}^{(t+1)}, \label{eq:position_PSOLPA} 
\end{equation}
\begin{equation}
\mathbf{J}_{v_z}^{(t+1)}\hspace{-0.5ex}=\mu_1\hspace{-0.25ex} \mathbf{Y}_1^{(t)}\hspace{-0.5ex}(\hspace{-0.35ex}\mathbf{J}_{p_\mathrm{best}}^{(t)}\hspace{-0.95ex}-\mathbf{J}_{p_z}^{(t)}\hspace{-0.25ex})\hspace{-0.25ex}+\hspace{-0.35ex}\mu_2\hspace{-0.25ex} \mathbf{Y}_2^{(t)}\hspace{-0.5ex}(\hspace{-0.35ex}\mathbf{J}_{p_{\mathrm{best}_z}}^{(t)}\hspace{-0.95ex}-\mathbf{J}_{p_z}^{(t)}\hspace{-0.25ex})\hspace{-0.15ex}+\hspace{-0.25ex}\mu_3^{(t)} \mathbf{J}_{v_z}^{(t)}\hspace{-0.25ex}. \label{eq:velocity_PSOLPA}  \vspace{-1ex}
\end{equation}
Finally, the personal and global best solutions for $z^{th}$ particle during $t^{th}$ iteration are obtained as follows: \vspace{-0.5ex}
\begin{equation}
\mathbf{J}_{p_{\mathrm{best}_z}\hspace{-0.4ex}}^{(t)}\hspace{-1ex} = \hspace{-1em} \argmax_{\hspace{-0.3ex}\mathbf{J}_{p_z}^{(t^*)}, \forall t^* = 0,1,\cdots, t} \hspace{-1ex}R_T(\mathbf{F}_1,\mathbf{B}_1,\mathbf{F}_2,\mathbf{B}_2,\mathbf{X}_{z}^{(t^*)\hspace{-0.25ex}}, \mathbf{P}_{\Phi_z}^{(t^*)}),   \vspace{-0.5ex}
\label{eq: personal_best_PSOLPA}
\end{equation} 
\begin{equation}
\mathbf{J}_{p_{\mathrm{best}}}^{(t)} \hspace{-1ex}= \hspace{-2ex} \argmax_{\hspace{-0.7ex}\mathbf{J}_{p_{\mathrm{best}_z}}^{(t)}, \forall z = 0,1,\cdots, Z\hspace{-1em}} \hspace{-1ex}R_T(\hspace{-0.2ex}\mathbf{F}_1\hspace{-0.2ex},\mathbf{B}_1\hspace{-0.2ex},\mathbf{F}_2\hspace{-0.2ex},\mathbf{B}_2\hspace{-0.2ex},\mathbf{X}_{\mathrm{best}_z}^{(t)\hspace{-0.25ex}}, \mathbf{P}_{\Phi_{\mathrm{best}_z}\hspace{-0.2ex}}^{(t)}),   \vspace{-0.5ex}
\label{eq: global_best_PSOLPA}
\end{equation} 
After $T$ iterations, we update $\mathbf{x}_o = \mathbf{X}_{\mathrm{best}}^{(T)}$ and $\mathbf{\Phi} = \diag(e^{(j\mathbf{P}_{\Phi_{\mathrm{best}}})})$. Algorithm 1 summarizes the proposed joint HBF, RIS phase shift, and deployment scheme.

\section{Numerical Results}

\begin{table}[!t]
	\caption{Simulation parameters.}
	\vspace{-1ex}
	\label{table_parameters}
	\centering
	\begin{tabular}{c|c }
		\hline
		\hline
	 	\# of transmit antennas             & $M_T= 8\times 8 = 64$                         \\ \hline
	 	\# of receive antennas              & $M_R= 8\times 8 = 64$                           \\ \hline
	 	Operating frequency                 & $28$ GHz                    \\ \hline
        \# of data streams                  & $N_S =2$ 
        \\ \hline
        Tx position                 & ($0,0,2$) m    
        \\ \hline
        User position                 & ($100,100,2$) m  
           \\ \hline
        Ceiling platform $x$-axis Range	              & $[40, 70]$ m   
        \\ \hline
        Ceiling platform $y$-axis Range	              & $[40, 70]$ m 
         \\ \hline
		\# of paths for ${\bf H}_l$  & $L=10$ for $l\in \left\{TI,IR\right\}$ \\ \hline
		Noise PSD                           & $-174$ dBm/Hz     \\ \hline
		Channel bandwidth                   & $10$ MHz          \\ \hline
		PSO \# of iterations                & $T=30$           \\ \hline
		PSO \# of particles                 & $N_\textrm{par}=10$           \\ \hline
		Path loss exponent     & $\eta=3.6$          
  \\ \hline
		Mean elevation  AoA/AoD                  &$\theta_{i}^{t}  =\theta_{i}^{r}=\ang{60}$,\\ \hline 
		Mean azimuth AoA/AoD                    & $\psi_{i}^{t}=\psi_{i}^{r}=\ang{120}$ 
   \\ \hline
		Angular spread                  &$\delta{i}^{\theta}  =\delta{i}^{\psi}  =\ang{10}$,
  
  \\ \hline
				\hline
	\end{tabular}
	\vspace{-3ex}
\end{table}

\begin{figure}[!t]
	\centering
	\includegraphics[width=1\columnwidth]{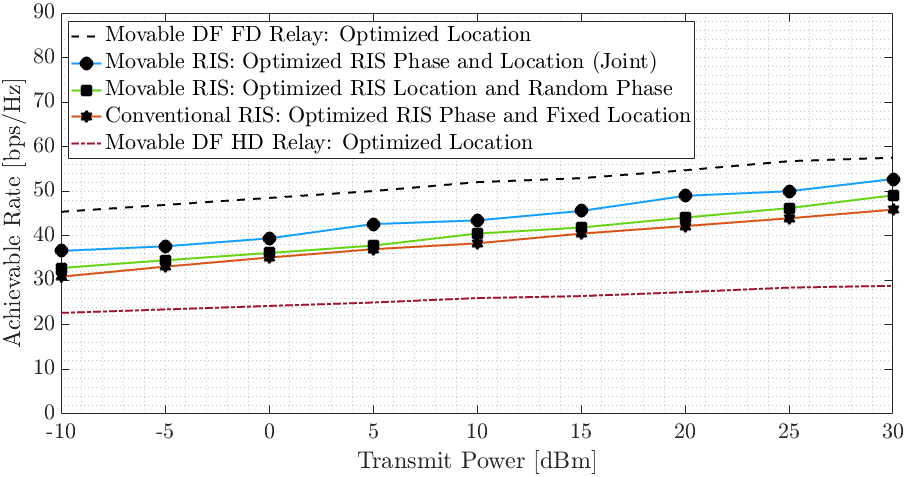}
	
	\caption{Achievable rate comparison of movable RIS and relay-aided AB-HPC systems under increasing  $P_T$.}
	\vspace{-3ex}
	\label{fig_Pt} 
\end{figure}	

In this section, we have analyzed the achievable rate performances of the proposed Spider RIS systems through computer simulations. For performance comparison, analyses have been conducted taking the traditional fixed RIS, full-duplex (FD) relay, and half-duplex (HD) relay-aided systems as references. The common system parameters for the provided simulations are presented in Table I. In our study, the total number of iterations is selected based on empirical testing to balance exploitation of the search space, ensuring convergence to a satisfactory solution with reasonable computational complexity.


In Fig. \ref{fig_Pt}, we investigate the effect of increasing $P_T$ on the achievable rate performances for the proposed Spider RIS architecture. To establish a benchmark, we consider movable FD and HD decode-and-forward (DF) relay-aided scenarios, serving as the upper and lower bounds to gauge the performance limits of our proposed system, respectively. According to the results obtained, it is observed that joint optimization of the RIS phase and position enhances the performance of RIS-aided communication systems with fixed location or random phase cases. Furthermore, even in the absence of phase optimization, Spider RIS systems outperform fixed-position RIS systems with optimized phase adjustment. This result indicates that mobility is an essential factor in improving the performance of RIS systems. Although the achievable rates of all systems increase as the transmission power increases, the system that reaches the performance of the movable FD relay-aided system is the Spider RIS system, where position and phase optimization are performed together.

\begin{figure}[!t]
	\centering
	\includegraphics[width=1\columnwidth]{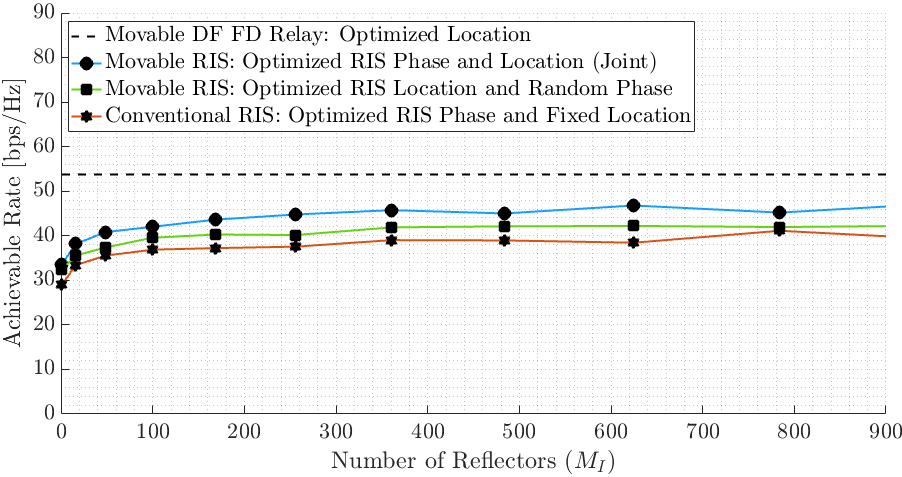} \vspace{-3ex}
	\caption{Achievable rate comparison of different phase adjustment methods in RIS-aided AB-HPC systems for varying $M_I$.}
	\vspace{-3ex}
	\label{fig_MI} 
\end{figure}

It is observed in Fig. \ref{fig_MI} that the increasing number of reflectors 
progressively enhances the performance of Spider RIS with that of movable FD relay-aided systems for $P_T=30$ dBm.
The increasing number of reflectors narrows the performance gap, bringing Spider RIS-aided systems in proximity to the achievable rate levels demonstrated by mobile FD relaying.
This proximity is pivotal, particularly concerning cost and energy efficiency, exploiting the inherently passive nature of RIS elements. In contrast to relays equipped with active elements, Spider RIS-aided systems provide comparable performance metrics at a notably lower cost and energy consumption. The cost-efficiency is attributed to RIS's passive nature and the use of low-cost, programmable elements. Unlike active relay systems, RISs require minimal power for operation, resulting in lower energy consumption and operational costs.
This emphasizes that RIS technology stands as a cost-effective and energy-efficient alternative within wireless communication systems through the strategic augmentation of reflector numbers. 

\begin{figure}[!t]
	\centering
	\includegraphics[width=1\columnwidth]{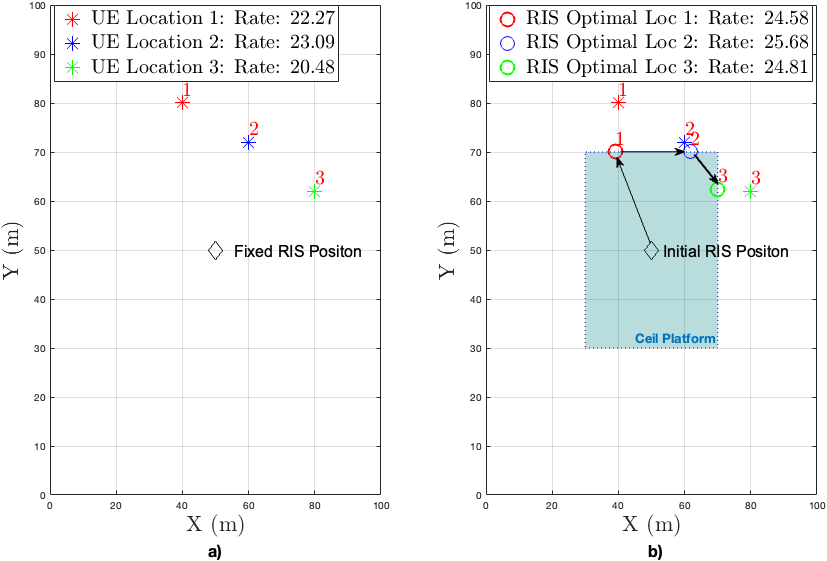}
	\vspace{-3ex}
	\caption{Achievable rate comparison of different phase adjustment methods in RIS-aided AB-HPC systems for varying $M_I$.}
	\vspace{-3ex}
	\label{fig_tra} 
\end{figure}

Fig. \ref{fig_tra} compares the performances of fixed-position RIS and mobile Spider RIS for three different locations of user equipment (UE). Fig. \ref{fig_tra}(a) shows the achievable rates provided by the fixed-position RIS according to UE locations, while Fig. \ref{fig_tra}(b) demonstrates the achievable rates obtained by Spider RIS after phase and position optimization according to three different locations of the UE.
In Fig. \ref{fig_tra}(a), with the fixed RIS position, the achievable rates for different UE locations vary between $20.48$ and $23.09$ bps/Hz, while in Fig. \ref{fig_tra}(b), it is observed that achievable rates improved between $24.58$ and $25.68$ bps/Hz with Spider RIS. This enhancement reflects that Spider RIS provides superior performance for each user location compared to fixed RIS with optimized phase adjustment. 
The observed results underline the superior performance of the Spider RIS in various user positions compared to the fixed RIS with optimized phase adjustment. The visual representation further illustrates Spider RIS's capability to dynamically transition from its initial position to optimal locations, all within a designated "Ceiling Platform". This not only exemplifies the inherent mobility of Spider RIS but also highlights its adaptability as it dynamically responds to various user locations. The overall implication of these findings is the transformative potential of mobile RIS technology, significantly raising the efficiency and adaptability benchmarks for wireless networks. This type of dynamic adaptability positions Spider RIS at the forefront of innovations that can seamlessly meet the diverse and evolving needs of wireless communications.

\section{Conclusions}
In this study, we have focused on the inherent benefits of Spider RIS, emphasizing its remarkable performance bolstered by strategic mobility and optimization capabilities.
Spider RIS offers superior performance than fixed RIS, providing more effective and dynamic transmission in changing user locations. The mobility and optimization strategy also allows Spider RIS to improve communication performance by increasing achievable rates. These findings highlight not only Spider RIS's adaptability but also its ability to react efficiently and dynamically to different user locations. The overall results of this study highlight the potential of Spider RIS to become a groundbreaking technology in wireless communications, indicating its potential to take a significant step toward increasing efficiency and adaptability in indoor and outdoor communication systems. The mechanical movement of the Spider RIS is one of the current drawbacks, and movement algorithms that will work more efficiently in the future need to be studied. Further research should also focus on the development of more sophisticated algorithms to jointly increase the efficiency of this mechanical movement and phase adjustment and further expand the potential of Spider RIS. 

\ifCLASSOPTIONcaptionsoff
\newpage
\fi
\bibliographystyle{IEEEtran}
\bibliography{bib_11_2023}
\balance

\end{document}